# Bidirectional Teleportation of a Pure EPR State by using GHZ states

Shima Hassanpour [*] . Monireh Houshmand [**]

**Abstract**

In the present paper, a novel bidirectional quantum teleportation protocol is proposed. By using entanglement swapping technique, two GHZ states are shared as a quantum channel between Alice and Bob as legitimate users. In this scheme, based on control-not operation, single-qubit measurement and appropriate unitary operations, two users can simultaneously transmit a pure EPR state to each other. While, in the previous protocols, the users can just teleport a single-qubit state to each other via more than four-qubit state. Therefore, the proposed scheme is economical compared with previous protocols.

## 1 Introduction

Quantum teleportation (QT) is one of the most remarkable protocols of quantum information theory [1] where an unknown quantum state can be transmitted to a receiver with the help of shared entanglement and some auxiliary classical communications. In 1993, Bennett et al., [2] proposed the original QT using an Einstein-Podolsky-Rosen (EPR) pair. After that, several quantum teleportation schemes have been proposed by using EPR pair, GHZ state and other entangled states as a quantum channel [3 − 8].

Bidirectional quantum controlled teleportation (BQCT) [9 − 15] as a field of QT has attracted a number of attentions in recent years. In 2013, Zha et al., [9] reported a BQCT scheme via five-qubit cluster state. In the same year, Li et al., [10] proposed a protocol by using a five-qubit composite GHZ-Bell state as a quantum channel. Then, Shukla et al., [17] showed the Li et al.'s scheme [10] is not a BQCT protocol.

In 2013, two protocols of bidirectional quantum controlled teleportation via six-qubit cluster state and six-qubit maximally-entangled state were introduced by Yan [11] and Sun et al., [12] respectively. One year later, Fu et al., [16] presented a bidirectional quantum teleportation (BQT) scheme by utilizing four-qubit cluster state as a quantum channel. In their scheme, two users can simultaneously exchange their single-qubit states applying Hadamard operation, defined unitary operations and Bell basis measurement. In that year, Chen [13] proposed a scheme by using five-qubit entangled state. In 2014, Duan et al., [14] presented a protocol via maximally seven-qubit entangled state as a quantum channel. Also, Duan et al. claimed utilizing seven-qubit entangled state improves the security of the scheme. In fact, the controller performs single-qubit measurement three times. In that year, a scheme via six-qubit entangled state was reported by Duan and Zha [15]. Also, they improved the security of their scheme by applying two single-qubit measurements.

In the best case of all the previous BQCT and BQT protocols [9 − 16], Alice and Bob can only teleport an arbitrary single-qubit state to each other with four-qubit state as a quantum channel. But, in this work, Alice and Bob as legitimate users can teleport a pure EPR state to each other via two GHZ states. This scheme can be prepared by controlled-not operation, single-qubit measurement and suitable unitary operations.

---

[*] Corresponding author, MS Student, Department of Electrical Engineering, Imam Reza International University, Mashhad, Iran Tel.: +98-882-179-9.
E-mail address: shimahassanpour@yahoo.com.

[**] Assistant Professor, Department of Electrical Engineering, Imam Reza International University, Mashhad, Iran
E-mail address: m_houshmand61@yahoo.com.

The remainder of this paper is organized as follows. The next section introduces the appropriate quantum channel. In Section 3, the proposed BQT protocol is studied in details. The comparison is given in Section 4. Finally, the conclusion is provided in Section 5.

## 2 Six-qubit entangled GHZ state as a quantum channel

Before generating the six-qubit entangled state as a quantum channel, we describe entanglement swapping technique [18] between two-GHZ states. The GHZ states are described as follows:

$$
\begin{aligned}
|\Psi_0\rangle &= \tfrac{1}{\sqrt{2}}(|000\rangle + |111\rangle), & |\Psi_4\rangle &= \tfrac{1}{\sqrt{2}}(|010\rangle + |101\rangle), \\
|\Psi_1\rangle &= \tfrac{1}{\sqrt{2}}(|000\rangle - |111\rangle), & |\Psi_5\rangle &= \tfrac{1}{\sqrt{2}}(|010\rangle - |101\rangle), \\
|\Psi_2\rangle &= \tfrac{1}{\sqrt{2}}(|100\rangle + |011\rangle), & |\Psi_6\rangle &= \tfrac{1}{\sqrt{2}}(|110\rangle + |001\rangle), \\
|\Psi_3\rangle &= \tfrac{1}{\sqrt{2}}(|100\rangle - |011\rangle), & |\Psi_7\rangle &= \tfrac{1}{\sqrt{2}}(|110\rangle - |001\rangle).
\end{aligned}
\quad (1)
$$

Suppose Alice and Bob share a couple of three-particle GHZ state ($|\Psi_0\rangle_{123}$ and $|\Psi_0\rangle_{456}$), where qubits 1, 3 and 5 are held by Alice and qubits 2, 4 and 6 are held by Bob. Now let Alice perform a GHZ basis measurement on her three qubits. Therefore, the states collapse to $|\Psi_0\rangle_{135}|\Psi_0\rangle_{246}$, $|\Psi_1\rangle_{135}|\Psi_1\rangle_{246}$, $|\Psi_6\rangle_{135}|\Psi_2\rangle_{246}$ and $|\Psi_7\rangle_{135}|\Psi_3\rangle_{246}$ with the same probability of 1/4:

$$
\begin{aligned}
|\Psi_0\rangle_{123} \otimes |\Psi_0\rangle_{456} &= \tfrac{1}{\sqrt{2}}(|000\rangle + |111\rangle)_{123} \otimes \tfrac{1}{\sqrt{2}}(|000\rangle + |111\rangle)_{456} \\
&= \tfrac{1}{2}(|000\rangle_{123}|000\rangle_{456} + |000\rangle_{123}|111\rangle_{456} + |111\rangle_{123}|000\rangle_{456} + |111\rangle_{123}|111\rangle_{456}) \\
&= \tfrac{1}{2}(|000\rangle_{135}|000\rangle_{246} + |001\rangle_{135}|011\rangle_{246} + |110\rangle_{135}|100\rangle_{246} + |111\rangle_{135}|111\rangle_{246}) \\
&= \tfrac{1}{2}(|\Psi_0\rangle_{135}|\Psi_0\rangle_{246} + |\Psi_1\rangle_{135}|\Psi_1\rangle_{246} + |\Psi_6\rangle_{135}|\Psi_2\rangle_{246} + |\Psi_7\rangle_{135}|\Psi_3\rangle_{246}).
\end{aligned}
\quad (2)
$$

In fact, the state after each user's measurement becomes one of the GHZ states that have maximally entangled state. According to Eq. (1), if Alice and Bob share other GHZ states, similar results can be achieved by utilizing this technique.

The quantum channel for implementing the bidirectional quantum teleportation in this paper is a six-qubit entangled state which is generated as follows:

$$
\begin{aligned}
|G\rangle_{a_1 b_1 b_2 a_2 a_3 b_3} &= \tfrac{1}{\sqrt{2}}(|000\rangle + |111\rangle)_{a_1 b_1 b_2} \otimes \tfrac{1}{\sqrt{2}}(|000\rangle + |111\rangle)_{a_2 a_3 b_3} \\
&= \tfrac{1}{2}(|000000\rangle + |000111\rangle + |111000\rangle + |111111\rangle)_{a_1 b_1 b_2 a_2 a_3 b_3}.
\end{aligned}
\quad (3)
$$

## 3 Description of the Proposed protocol

In this protocol, Alice and Bob as a two legitimate users want to teleport a two-qubit entangled state to each other. Suppose Alice and Bob have the pure EPR state, which are described as Eq. (4),

$$
\begin{aligned}
|\emptyset\rangle_{A_1 A_2} &= \alpha_0|00\rangle + \alpha_1|11\rangle, \\
|\emptyset\rangle_{B_1 B_2} &= \beta_0|00\rangle + \beta_1|11\rangle,
\end{aligned}
\quad (4)
$$

where $|\alpha_0|^2 + |\alpha_1|^2 = 1$ and $|\beta_0|^2 + |\beta_1|^2 = 1$. This scheme consists of the following steps:

Step1. Assume that Alice and Bob share a two-GHZ entangled state as Eq. (3), where the qubits $a_1, a_2$ and $a_3$ belong to Alice and qubits $b_1, b_2$ and $b_3$ belong to Bob. The state of the whole system can be expressed as Eq. (5),

$$
|\Psi\rangle_{a_1 b_1 b_2 a_2 a_3 b_3 A_1 A_2 B_1 B_2} = |G\rangle_{a_1 b_1 b_2 a_2 a_3 b_3} \otimes |\emptyset\rangle_{A_1 A_2} \otimes |\emptyset\rangle_{B_1 B_2}. \quad (5)
$$

Step2. In this step, Alice and Bob make a controlled-not operation with qubits $A_1$ and $B_1$ as control qubits and qubits $a_1$ and $b_3$ as target respectively. The state will be the form of Eq. (6).

$$
\begin{aligned}
|\Psi'\rangle_{a_1 b_1 b_2 a_2 a_3 b_3 A_1 A_2 B_1 B_2} \\
= \tfrac{1}{2}[&(|000000\rangle + |000111\rangle + |111000\rangle + |111111\rangle)_{a_1 b_1 b_2 a_2 a_3 b_3} \alpha_0 \beta_0 |0000\rangle_{A_1 A_2 B_1 B_2} \\
+&(|000001\rangle + |000110\rangle + |111001\rangle + |111110\rangle)_{a_1 b_1 b_2 a_2 a_3 b_3} \alpha_0 \beta_1 |0011\rangle_{A_1 A_2 B_1 B_2} \\
+&(|100000\rangle + |100111\rangle + |011000\rangle + |011111\rangle)_{a_1 b_1 b_2 a_2 a_3 b_3} \alpha_1 \beta_0 |1100\rangle_{A_1 A_2 B_1 B_2} \\
+&(|100001\rangle + |100110\rangle + |011001\rangle + |011110\rangle)_{a_1 b_1 b_2 a_2 a_3 b_3} \alpha_1 \beta_1 |1111\rangle_{A_1 A_2 B_1 B_2}].
\end{aligned}
\quad (6)
$$

Step3. Alice and Bob perform a single-qubit measurement in the Z-basis on qubits $a_1$ and $b_3$ and the X-basis measurement on qubits $A_2$ and $B_2$ respectively. According to Table I, the remaining particles may collapse into one of the 16 possible state with the same probability.

TABLE I. THE MEASUREMENT RESULTS OF USERS AND THE CORRESPONDING COLLAPSED STATE

| Alice's results | Bob's results | The collapsed state of qubits $b_1, b_2, a_2, a_3, A_1, B_1$ |
|---|---|---|
| $\|0\rangle_{a_1}\|+\rangle_{A_2}$ | $\|0\rangle_{b_3}\|+\rangle_{B_2}$ | $\alpha_0\beta_0\|000000\rangle + \alpha_0\beta_1\|010011\rangle + \alpha_1\beta_0\|101100\rangle + \alpha_1\beta_1\|111111\rangle$ |
| $\|0\rangle_{a_1}\|+\rangle_{A_2}$ | $\|0\rangle_{b_3}\|-\rangle_{B_2}$ | $\alpha_0\beta_0\|000000\rangle - \alpha_0\beta_1\|010011\rangle + \alpha_1\beta_0\|101100\rangle - \alpha_1\beta_1\|111111\rangle$ |
| $\|0\rangle_{a_1}\|-\rangle_{A_2}$ | $\|0\rangle_{b_3}\|+\rangle_{B_2}$ | $\alpha_0\beta_0\|000000\rangle + \alpha_0\beta_1\|010011\rangle - \alpha_1\beta_0\|101100\rangle - \alpha_1\beta_1\|111111\rangle$ |
| $\|0\rangle_{a_1}\|-\rangle_{A_2}$ | $\|0\rangle_{b_3}\|-\rangle_{B_2}$ | $\alpha_0\beta_0\|000000\rangle - \alpha_0\beta_1\|010011\rangle - \alpha_1\beta_0\|101100\rangle + \alpha_1\beta_1\|111111\rangle$ |
| $\|0\rangle_{a_1}\|+\rangle_{A_2}$ | $\|1\rangle_{b_3}\|+\rangle_{B_2}$ | $\alpha_0\beta_0\|000011\rangle + \alpha_0\beta_1\|010000\rangle + \alpha_1\beta_0\|101111\rangle + \alpha_1\beta_1\|111100\rangle$ |
| $\|0\rangle_{a_1}\|+\rangle_{A_2}$ | $\|1\rangle_{b_3}\|-\rangle_{B_2}$ | $\alpha_0\beta_0\|000011\rangle - \alpha_0\beta_1\|010000\rangle + \alpha_1\beta_0\|101111\rangle - \alpha_1\beta_1\|111100\rangle$ |
| $\|0\rangle_{a_1}\|-\rangle_{A_2}$ | $\|1\rangle_{b_3}\|+\rangle_{B_2}$ | $\alpha_0\beta_0\|000011\rangle + \alpha_0\beta_1\|010000\rangle - \alpha_1\beta_0\|101111\rangle - \alpha_1\beta_1\|111100\rangle$ |
| $\|0\rangle_{a_1}\|-\rangle_{A_2}$ | $\|1\rangle_{b_3}\|-\rangle_{B_2}$ | $\alpha_0\beta_0\|000011\rangle - \alpha_0\beta_1\|010000\rangle - \alpha_1\beta_0\|101111\rangle + \alpha_1\beta_1\|111100\rangle$ |
| $\|1\rangle_{a_1}\|+\rangle_{A_2}$ | $\|0\rangle_{b_3}\|+\rangle_{B_2}$ | $\alpha_0\beta_0\|001100\rangle + \alpha_0\beta_1\|011111\rangle + \alpha_1\beta_0\|100000\rangle + \alpha_1\beta_1\|110011\rangle$ |
| $\|1\rangle_{a_1}\|+\rangle_{A_2}$ | $\|0\rangle_{b_3}\|-\rangle_{B_2}$ | $\alpha_0\beta_0\|001100\rangle - \alpha_0\beta_1\|011111\rangle + \alpha_1\beta_0\|100000\rangle - \alpha_1\beta_1\|110011\rangle$ |
| $\|1\rangle_{a_1}\|-\rangle_{A_2}$ | $\|0\rangle_{b_3}\|+\rangle_{B_2}$ | $\alpha_0\beta_0\|001100\rangle + \alpha_0\beta_1\|011111\rangle - \alpha_1\beta_0\|100000\rangle - \alpha_1\beta_1\|110011\rangle$ |
| $\|1\rangle_{a_1}\|-\rangle_{A_2}$ | $\|0\rangle_{b_3}\|-\rangle_{B_2}$ | $\alpha_0\beta_0\|001100\rangle - \alpha_0\beta_1\|011111\rangle - \alpha_1\beta_0\|100000\rangle + \alpha_1\beta_1\|110011\rangle$ |
| $\|1\rangle_{a_1}\|+\rangle_{A_2}$ | $\|1\rangle_{b_3}\|+\rangle_{B_2}$ | $\alpha_0\beta_0\|001111\rangle + \alpha_0\beta_1\|011100\rangle + \alpha_1\beta_0\|100011\rangle + \alpha_1\beta_1\|110000\rangle$ |
| $\|1\rangle_{a_1}\|+\rangle_{A_2}$ | $\|1\rangle_{b_3}\|-\rangle_{B_2}$ | $\alpha_0\beta_0\|001111\rangle - \alpha_0\beta_1\|011100\rangle + \alpha_1\beta_0\|100011\rangle - \alpha_1\beta_1\|110000\rangle$ |
| $\|1\rangle_{a_1}\|-\rangle_{A_2}$ | $\|1\rangle_{b_3}\|+\rangle_{B_2}$ | $\alpha_0\beta_0\|001111\rangle + \alpha_0\beta_1\|011100\rangle - \alpha_1\beta_0\|100011\rangle - \alpha_1\beta_1\|110000\rangle$ |
| $\|1\rangle_{a_1}\|-\rangle_{A_2}$ | $\|1\rangle_{b_3}\|-\rangle_{B_2}$ | $\alpha_0\beta_0\|001111\rangle - \alpha_0\beta_1\|011100\rangle - \alpha_1\beta_0\|100011\rangle + \alpha_1\beta_1\|110000\rangle$ |

Step4. After Alice and Bob tell their results to each other, they carry out X-basis measurement on their qubits $A_1$ and $B_1$ and announce their results. Therefore, by applying appropriate operations as we can see in Table II, each one can reconstruct the other's two-qubit entangled state. Also, if Alice (Bob) does not announce his results to Bob (Alice), Bob (Alice) cannot reconstruct the other's two-qubit entangled state. As an example, if Alice's measurements result in the first step is $0_{a_1} +_{A_2}$ and Bob's measurements result is $0_{b_3} +_{B_2}$, the state of the remaining particles collapse into the state as Eq. (7),

$$|\Omega\rangle_{b_1 b_2 a_2 a_3 A_1 B_1}$$
$$= (\alpha_0\beta_0|000000\rangle + \alpha_0\beta_1|010011\rangle + \alpha_1\beta_0|101100\rangle + \alpha_1\beta_1|111111\rangle)_{b_1 b_2 a_2 a_3 A_1 B_1}. \qquad (7)$$

After Alice and Bob apply another single-qubit measurement, the state will change into the following state,

$$\begin{aligned}
|\Omega\rangle_{A_1B_1b_1a_2b_2a_3} = &|+\rangle_{A_1}|+\rangle_{B_1}(\alpha_0\beta_0|0000\rangle + \alpha_0\beta_1|0011\rangle + \alpha_1\beta_0|1100\rangle + \alpha_1\beta_1|1111\rangle)_{b_1b_2a_2a_3} \\
&+|+\rangle_{A_1}|-\rangle_{B_1}(\alpha_0\beta_0|0000\rangle - \alpha_0\beta_1|0011\rangle + \alpha_1\beta_0|1100\rangle - \alpha_1\beta_1|1111\rangle)_{b_1b_2a_2a_3} \\
&+|-\rangle_{A_1}|+\rangle_{B_1}(\alpha_0\beta_0|0000\rangle + \alpha_0\beta_1|0011\rangle - \alpha_1\beta_0|1100\rangle - \alpha_1\beta_1|1111\rangle)_{b_1b_2a_2a_3} \\
&+|-\rangle_{A_1}|-\rangle_{B_1}(\alpha_0\beta_0|0000\rangle - \alpha_0\beta_1|0011\rangle - \alpha_1\beta_0|1100\rangle + \alpha_1\beta_1|1111\rangle)_{b_1b_2a_2a_3} \\
= &|+\rangle_{A_1}|+\rangle_{B_1}(\alpha_0|00\rangle + \alpha_1|11\rangle)_{b_1b_2}(\beta_0|00\rangle + \beta_1|11\rangle)_{a_2a_3} \\
&+|+\rangle_{A_1}|-\rangle_{B_1}(\alpha_0|00\rangle + \alpha_1|11\rangle)_{b_1b_2}(\beta_0|00\rangle - \beta_1|11\rangle)_{a_2a_3} \\
&+|-\rangle_{A_1}|+\rangle_{B_1}(\alpha_0|00\rangle - \alpha_1|11\rangle)_{b_1b_2}(\beta_0|00\rangle + \beta_1|11\rangle)_{a_2a_3} \\
&+|-\rangle_{A_1}|-\rangle_{B_1}(\alpha_0|00\rangle - \alpha_1|11\rangle)_{b_1b_2}(\beta_0|00\rangle - \beta_1|11\rangle)_{a_2a_3}.
\end{aligned} \quad (8)$$

Now, each legitimate user can reconstruct the two-qubit entangled state by applying suitable unitary operation as can be seen in Table II. Thus, the bidirectional teleportation is successfully finished.

TABLE II. RELATION BETWEEN THE MEASUREMENT RESULTS AND APPROPRIATE UNITARY OPERATIONS

| Alice's measurement result | Bob's measurement result | Bob's operation | Alice's operation |
|---|---|---|---|
| $|+\rangle_{A_1}$ | $|+\rangle_{B_1}$ | $I_{b_1} \otimes I_{b_2}$ | $I_{a_2} \otimes I_{a_3}$ |
| $|+\rangle_{A_1}$ | $|-\rangle_{B_1}$ | $\sigma^z_{b_1} \otimes \sigma^z_{b_2}$ | $I_{a_2} \otimes \sigma^z_{a_3}$ |
| $|-\rangle_{A_1}$ | $|+\rangle_{B_1}$ | $I_{b_1} \otimes \sigma^z_{b_2}$ | $\sigma^z_{a_2} \otimes \sigma^z_{a_3}$ |
| $|-\rangle_{A_1}$ | $|-\rangle_{B_1}$ | $\sigma^z_{b_1} \otimes I_{b_2}$ | $\sigma^z_{a_2} \otimes I_{a_3}$ |

## 4 Comparison

The most important advantage of the proposed protocol is that two users can simultaneously teleport a pure EPR pair two each other via two GHZ states as a quantum channel. While, in the best case of previous schemes [9 − 16] the users can just teleport a single-qubit state by applying four-qubit state. Therefore, the previous protocols need to be run two times in order to teleport two-qubit state. Unlike previous works which users apply two-qubit measurements, in the proposed scheme, the users utilize single-qubit measurements which are more efficient than two-qubit measurements [20].

Also, we use the idea of entanglement swapping technique for sharing two GHZ states as a quantum channel. Therefore, the quantum channel in the proposed scheme is easier to prepare in the experiment compared to the previous works which their quantum channel are cluster or brown state [19]. In the proposed scheme, the entanglement must be preserved between three qubits, while maintaining entanglement between more than three qubits is more complicated. The details of comparison between the proposed protocol and previous works are given in Table III.

TABLE III. COMPARISON BETWEEN PROPOSED BQT PROTOCOL WITH PREVIOUS WORKS

| Protocol | Type of protocol | Quantum channel | teleported Quantum state | Type of measurement |
|---|---|---|---|---|
| Protocol [16] | BQT | Cluster$_4$ | Single-qubit state | Two-qubit measurement |
| Protocol [9] | CBQT | Cluster$_5$ | Single-qubit state | Single-qubit measurement |
| Protocol [11] | CBQT | Cluster$_6$ | Single-qubit state | Single-qubit measurement |
| Protocol [12] | CBQT | Six-qubit | Single-qubit state | Two-qubit measurement |
| Protocol [13] | CBQT | Brown$_5$ | Single-qubit state | Two-qubit measurement |
| Protocol [14] | CBQT | Seven-qubit | Single-qubit state | Two-qubit measurement |
| Protocol [15] | CBQT | Six-qubit | Single-qubit state | Two-qubit measurement |
| **Proposed protocol** | **BQT** | **GHZ$_6$** | **Pure EPR state** | **Single-qubit measurement** |

## 5 Conclusions

In this paper, a new bidirectional quantum teleportation scheme via entanglement swapping is presented, where the quantum channel are two GHZ states. In the proposed scheme, Alice and Bob can teleport a pure EPR state to each other simultaneously with single-qubit measurement, controlled-not operation and appropriate unitary operations. Also, if one user does not cooperate, the receiver cannot reconstruct the original state.

Unlike other BQCT schemes in which the users can just teleport a single-qubit state, in the proposed protocol, Alice and Bob teleport a two-qubit state to each other. Indeed, in the the best case of previous works, single-qubit state is teleported via four-qubit state as a quantum channel. But, in the proposed scheme, a pure EPR state is transmitted by applying two GHZ states as a quantum channel. Therefore, the present protocol is economical compared with other schemes in this field.

We hope the proposed BQT protocol will be extended to a bidirectional quantum controlled teleportation protocol in near future which Alice and Bob may teleport more than two qubits to each other under the control of Charlie.

**References**

[1]. M. A. Nielsen, I. L. Chuang, Cambridge University Press, Cambridge, 2002.
[2]. C. H. Bennet, G. Brassard, C. Crepeau, R. Jozsa, A. Peres, W. K. Wooters, Phys. Rev. Lett. 70 (1993) 1895.
[3]. G. Rigolin, Phys. Rev. A 71 (2005) 032303.
[4]. W. B. Cardoso, Int. J. Theor. Phys. 47 (2008) 977.
[5]. B. S. Shi, A. Tomita, Phys. Lett. A 296 (2002).
[6]. K. Yang, L. Huang, W. Yang, F. Song, Int. J. Theor. Phys. 48 (2009) 516.
[7]. D. Tian, Y. Tao, M. Qin, Science in China Series G: Physics, Mechanics and Astronomy. 51 (2008) 1523.
[8]. S. Q. Tang , C. J. Shan, X. X. Zhang, Int. J. Theor. Phys. 49 (2010) 1899.
[9]. X. W. Zha, Z. C. Zou, J. X. Qi, H.Y. Song, Int J Theor Phys. 52 (2013) 1740.
[10]. Y. H. Li, L. P. Nie, Int J Theor Phys. 52 (2013) 1630.
[11]. A. Yan, Int J Theor Phys. 52 (2013) 3870.
[12]. X. M. Sun, X. W. Zha, Acta. Photonica. Sinica. 48 (2013) 1052.
[13]. H. Z. Fu, X. L. Tian, Y. Hu, Int. J. Theor. Phys. 53 (2014) 1840.
[14]. Y. Chen, Int J Theor Phys. 53 (2014) 1454.
[15]. Y. J. Duan, X.W. Zha, X.M. Sun, J.F. Xia, Int J Theor Phys. 53 (2014) 2697.
[16]. Y. J. Duan, X.W. Zha, Int J Theor Phys, April 2014.
[17]. C. Shukla, A. Baerjee, A. Pathak, Int J Theor Phys. 52 (2013) 3790.
[18]. S. Hassanpour, M. Houshmand, Quantum Inf. Process., ( Accepted for publication) to be published.
[19]. M. Lucamarini, S. Mancini, Phys. Rev. Lett. 94 (2005) 140501.
[20]. F. G. Deng, X. H. Li, C. Y. Li, P. Zhou, H. Y. Zhou, Eur. Phys. J. D. 39 (2006) 459.